\def\comment#1{}
\documentstyle[pra,epsf,aps,twocolumn]{revtex}
\begin{document}
 \title{Proper Dirac Quantization of Free Particle on
$D$-Dimensional Sphere
}
\author{
Hagen KLEINERT and
Sergei V. SHABANOV\thanks{\noindent
Alexander von Humboldt fellow;
on leave from Laboratory of Theoretical
Physics, JINR, Dubna, Russia.}
\thanks{
Email:
kleinert@physik.fu-berlin.de;
shabanov@physik.fu-berlin.de;
 URL: http://www.physik.fu-berlin.de/\~{}kleinert. Phone/Fax: 0049/30/8383034
}}
\address{Institut f\"ur Theoretische Physik,
Freie Universit\"at Berlin, Arnimallee 14, 14195 Berlin, Germany}
\maketitle
\begin{abstract}
We show that an unambiguous and correct quantization of
the second-class constrained
system of a free particle on a sphere in $D$ dimensions
is possible only by converting the constraints
to abelian gauge constraints, which are
of first class in Dirac's classification scheme.
The energy spectrum is equal to that of a pure Laplace-Beltrami operator
with no additional constant arising from the curvature
of the sphere.
A quantization of Dirac's modified
Poisson brackets
for second-class constraints
is also possible and unique,
but must be rejected since the resulting
energy spectrum is physically incorrect.
\end{abstract}
{}~\\
{\bf 1}. Quantization of a free point particle in
curved space is a long-standing
and controversial problem in quantum mechanics. Dirac has emphasized
that
canonical quantization rules are consistent only in a Cartesian
reference
frame \cite{dirac1}.
Attempts to generalize these rules
to
curved space run into the
notorious
{\em operator-ordering problem\/}
of momentum and coordinates,
 making the Hamiltonian operator non-unique.
Podolsky\cite{POD}
avoided this problem by
{\em postulating\/}
that the Laplacian in the free Schr\"odinger operator
$H=-\hbar ^2  \Delta/2$
should be replaced by
the Laplace-Beltrami operator $ \Delta_{\rm LB}= {g}^{-1/2}\partial _\mu
g^{1/2}g^{\mu \nu}\partial _ \nu $,
where $\partial _\mu=\partial /\partial q^\mu$ are the partial derivatives with
respect to
the $D$-dimensional curved-space coordinates, and $g$ is the determinant of the
metric $g_{\mu \nu}(q)$.
This postulate has generally been
accepted as being
correct
since it yields, for a $D$-dimensional sphere of radius
$R$
embedded in a $D+1$ -dimensional
Cartesian space
with coordinates $x^i$, an energy
$ \hat L_a^2/2R^2$.
Here $\hat L_a =-i\hat p_i (
L_a)_{ij} x_j$ with $\hat p_i=-i\hbar\partial /\partial x_i$
are the unique quantum-mechanical
differential operator representation
of
the $D(D+1)/2$ generators $L_a$
of the rotation
group  SO($D+1$)
in flat space. If we take
$a$ to label the index pairs $ij$ with
$i\le j$,  then $L_a$ are the matrices
 $(L_{ij})_{kl}=i(
\delta_{ik}\delta_{jl}-\delta_{ik}\delta_{jl})$,
whose
nonzero
commutation rules are
\begin{equation}
[L_{ij},L_{ik}]=-i L_{jk}\ \ \   ({\rm no\ sum\ over\ }i) .
\label{3}\end{equation}
The canonical commutation rules $[\hat p_i,\hat x_j]=-i\hbar  \delta_{ij}$
transfer the Lie algebra (\ref{3})
to the operators $\hat L_{ij}/\hbar$.
The square of the total angular momentum
$\hat L_a^2$ is the Casimir operator of the orthogonal group
SO(D+1),
with the eigenvalues $l(l+D-1)$, $~l=0,1,2,\dots~$ \cite{PI}.
Thus, by quantizing
classical
angular momentum rather than canonical variables
(i.e., by putting hats on $L$'s rather than on $p$'s)
operator-ordering problems are avoided,
making this energy spectrum
the most plausible one \cite{LL}.

Doubts on the correctness of this spectrum have been raised
by DeWitt\cite{dewitt} in his first attempt to
quantize the system by a straightforward generalization
of Feynman's time-sliced path integral
to curvilinear coordinates.
He found
an extra energy
proportional to the Riemannian
scalar curvature $\bar R$ of
in the Schr\"odinger operator:
\begin{equation}
H=-\frac{\hbar ^2}{2}\Delta_{\rm LB} +\alpha \hbar ^2\bar R\ ,
\label{1}
\end{equation}
with
a proportionality constant $ \alpha=1/24$.
Various successors have presented
modifications of
DeWitt's procedure
leading to
other proportionality factors
$ \alpha=1/12$ \cite{cheng}
and $ \alpha=1/8$ \cite{dekker}.

For the above sphere, the extra term produces an extra constant energy
equal to $ \alpha \hbar ^2 D(D-1)/R^2$
which would contradict the previous results.
At fist sight, this contradiction seems to be
rather irrelevant.
Common  experiments
are only capable
of detecting {\em energy differences\/},
 in which this constant drops out.
Cosmology, however, is sensitive to an additive constant,
which  would change the gravitational energy of interstellar rotating
two-atomic gases.
Apart from such somewhat esoteric applications, a correct quantization of
this simplest
non-Euclidean system is certainly of fundamental
theoretical interest.

It is therefore important to find physical systems for which the
$\bar R$-term
is not a constant,  so that it can distort the energy spectrum
of the Schr\"odinger operator
(\ref{1}). Such a system has recently been found, although in a somewhat
indirect way.
When solving the
path integral of the hydrogen atom,
a two-step nonholonomic mapping
is needed which passes through an intermediate non-Euclidean space
\cite{kleinert,PI2}.
The existence of an extra $\bar R$-term in (\ref{1})
would modify
significantly the
known level spacings
in the hydrogen spectrum.
Thus experiment
eliminates an
extra curvature scalar,
in agreement
with Podolsky
and with the quantization of angular momentum.
The successful procedure
was generalized
into
a simple
{\em nonholonomic mapping principle\/},
which allows to map classical and quantum-mechanical
laws in flat space into
correct laws in curved space \cite{kleinert,PI2,qep}.

The absence of an $\bar R$-term in (\ref{1}) is therefore
a crucial test for any quantum theory in
curved spaces.
It is
the  purpose of this note to show that
this absence
can be established
for a sphere
also within
the operator approach to quantum mechanics,
after suitably preparing the description
for an application
of
Dirac's
theory of constrained systems.

This application is,
however, not straight-forward.
A
$D$-dimensional sphere is most simply described
by
embedding it into a $D+1$ -dimensional Cartesian $x$-space
via
a constraint
$x^2-R^2=0$.
Within Dirac's classification scheme
\cite{dirac2}, such a
 constraint
 is of second class.
We shall demonstrate that
Dirac's quantization rules for such systems
produce
wrong energy levels.
A correct quantization becomes possible
only by making use of
a recently developed
conversion \cite{fradkin} of second-class constraints to
first-class constraints, which instead of configuration space
restrict
the quantum-mechanical {\em Hilbert space\/}
\cite{dirac2}.
This conversion requires an extension
of the phase space of
the initial Cartesian system
to a larger
auxiliary Cartesian phase space, where
the correct quantization is known. The operators representing the
first-class constraints are generators
of gauge transformations, and the physical
states are all found by going into
the gauge invariant subspace
of the Hilbert space.

Constraints associated with gauge symmetry
 have first been
mastered in quantum electrodynamics (Coulomb's law),
and are now a standard tool in
the quantization of gauge theories \cite{jackiw}.

{\bf 2}. We begin by quantizing
 the system with the help of
Dirac's theory for second-class constraints.
A free point particle
with a Hamiltonian $H=p^2/2$
moving in
a flat Euclidean $D+1$-dimensional
is restricted to the surface of a $D$-dimensional sphere
by the primary constraint in configuration space
\cite{dirac2}
\begin{equation}
\varphi _1=x^2-R^2=0\ .
\label{5}
\end{equation}
The dynamical consistency condition $\dot{\varphi}_1 =
\{H,\varphi_1\} =0$ leads to an additional secondary  constraint
in phase space
\begin{equation}
\varphi _2=(x,p)=0\ .
\label{5p}
\label{}\end{equation}
It expresses the fact that a motion on a sphere
has no radial component.
Although the canonical variables are Cartesian,
the canonical quantization is not applicable,
since
the
constraints (\ref{5}), (\ref{5p}) cannot be enforced for
 the associated operators --- the conditions
$\hat{\varphi} _1=\hat{\varphi}_2=0$ would be
in conflict with the commutation
relation $[\hat{\varphi}_1,\hat{\varphi}_2]=2i\hbar
\hat{x}^2=2i\hbar R^2\neq 0$.
To resolve this difficulty, Dirac replaced the Poisson
symplectic
structure $\{x_i,p_j\}=\delta _{ij}$
by
the so-called {\em Dirac brackets\/}
\begin{equation}
\{A,B\}_D=\{A,B\}-\{A,\varphi _a\}[\Delta^{-1}]^{ab}
\{\varphi _b,B\}\ ,
\label{6}
\end{equation}
where
$\Delta
_{ab}$ is a matrix
$ \Delta_{ab}=\{\varphi _a,\varphi _b\}$
formed from all primary and secondary
constraints $\varphi _a=0$. This matrix is assumed to
be non-degenerate, a defining
property of
second-class
constraints \cite{dirac2}).
The Dirac brackets
provide us
with an antisymmetric operation which is
 distributive and associative, i.e.,
which satisfies
Leibnitz rule and Jacobi identity, thus
forming a
symplectic structure which is as good as Poisson's.

The removal of the inconsistency is ensured
by the automatic property
\begin{equation}
\{A,\varphi _a\}_D=0\ ,\ \ \ \
{\rm   for\ any}\ \  A \ \ {\rm and}\ \  \varphi _a \ .
 \label{prop}\end{equation}

Hamiltonian
equations of motion generated by the Dirac bracket $\dot{A}=\{A,H\}_D$
coincide with those generated by the original Poisson bracket $\{\ ,\ \}$
on the surface of constraints $\varphi _a =0$.
Thus, Dirac brackets produce
classically
the same  equations of motion as Poisson brackets.
Quantization proceeds by the replacement
$ \{\ ,\ \}_D\rightarrow -i[\ ,\ ]/\hbar $.
The property (\ref{prop})
allows us to replace the constraints by operator
equations
$\hat{\varphi}_a=0$ without the earlier contradictions.
Moreover, since
the constraint operators $\hat{\varphi}_a$ commute with any
other operator, they can be given any $c$-number value,
for instance zero.
For the surface of the sphere,
the quantized Dirac brackets (\ref{6}) read
\begin{eqnarray}
\left[\hat{x}_i,\hat{x}_j\right]&= &0\ ; \label{7} \\
\left[\hat{x}_j,\hat{p}_k\right]&= & i\hbar\left(\delta
_{jk}-\frac{\hat{x}_j\hat{x}_k}{\hat{x}^2}\right)\ ; \label{8} \\
\left[\hat{p}_j,\hat{p}_k\right]&=&i\hbar
\frac{1}{\hat{x}^2}(\hat{p}_j\hat{x}_k- \hat{p}_k\hat{x}_j)\ . \label{9}
\end{eqnarray}
The operator ordering problem occurring in the right-hand side of (\ref{9})
is uniquely resolved under the condition that the algebra
(\ref{7})--(\ref{9}) satisfies the Jacobi identity.
The
operator $\hat{x}^2$ commutes with all canonical operators
and is therefore a $c$-number, which can be set equal $R^2$.

To find the spectrum of the Hamiltonian $\hat{H}=\hat{p}^2/2$, we make use
of the identity
\begin{equation}
\delta _{ij}=-\frac{(L_a\hat{x})_i
(L_a\hat{x})_j}{\hat{x}^2}+
\frac{\hat{x}_i\hat{x}_j}{\hat{x}^2}\ .
\label{10}
\end{equation}
Inserting this into $\hat{p}^2=\hat{p}_i\delta
_{ij}\hat{p}_j$, we obtain  the Hamiltonian
\begin{equation}
\hat{H}=\frac{1}{2R^2}\hat{L}^2_a+\frac{1}{2}\hat{p}^\dagger_r\hat{p}_r\ ,
\label{11}
\end{equation}
where $\hat{L}_a = -i\hat{p}_i(L_a)_{ij}\hat{x}_j$ as before and
$\hat{p}_r=(\hat{x},\hat{p})/R$ is a radial momentum operator.
Note that the operator (\ref{11})
is determined without operator-ordering ambiguities.
The identity
(\ref{10}) can equally well be inserted to the left and the right
of the momentum operators $\hat p_i\hat p_j$,
with a unique result.

The radial operator $\hat{p}_r$ commutes with
all canonical operators and is therefore some complex
number $c$. It has necessarily an
imaginary part, due to
the obvious relation
$\hat{p}_r-\hat{p}^\dagger_r=i\hbar D/R$ implied by
(\ref{8}). Thus we may decompose
$\hat{p}_r=c=i\hbar D/2R+c_r$, where $c_r$ is an arbitrary real
number.
The constant $c_r$ is determined by expressing
$\hat p_r $ in terms of the operator
of the second constraint (\ref{5p}).
The ordering of $\hat p$ and $\hat x $ in it is undetermined,
so that
$\hat \varphi_2$ may be written as $(\hat p_r+\hat p_r)/2
+i\hbar  \gamma$
with an arbitrary imaginary part $ \gamma$.
Inserting the above constant for $\hat p_r$ and setting
$\hat \varphi_2$ equal to zero is solved uniquely by
$ \gamma=0$ and $c_r=0$.

It is now easy to
find the spectrum of the first term in the Hamiltonian operator
(\ref{11}).
The
modified canonical rules
(\ref{7})--(\ref{9})
transfer the Lie algebra (\ref{3})
to the operators
$\hat L_a=-i\hat p L_a  \hat x$
in just the same way as
 ordinary canonical commutation rules,
so that the
energy levels of angular momentum $l$ are
\begin{equation}
E_l=\frac{\hbar ^2}{2R^2}l(l+D-1)+\frac{1}{8R^2}\hbar ^2D^2\ .
\label{12}
\end{equation}
The additional constant energy
has a different $D$-dependence than the
$\bar R$-term proposed by DeWitt and others \cite{cheng,dekker}.
In particular, it
is nonzero even for a particle on a circle ($D=2$),
where nobody ever expected such a term.

Thus the modified
symplectic structure proposed by Dirac, although
esthetically
appealing and yielding a unique result
 for a particle on the surface of a
sphere,
must be rejected on physical grounds
as contradicting the established quantization via angular momentum
operators
and nonholonomic mapping principle.

{\bf 3}. In second-class constrained systems, not every phase
space variable can be made an operator.
Dirac's
new symplectic structure (\ref{6}) represented by
(\ref{7})--(\ref{9})
accounts for this fact via its degeneracy
in the embedding phase space.
This permits
the
canonical variables of fluctuations
transverse to the manifold on which the
particle moves to remain $c$-numbers.
There is, however, a defect in Dirac's procedure:
Although
physical
excitations of
the transverse degree of freedom are eliminated,
the system maintains a memory of the forbidden motion
by a
nonzero
$c$-number valued
transverse energy $\hat{p}_r^\dagger\hat{p}_r/2$.
{}From the physical point of view, the existence
of such a memory must be rejected.
 After all, the embedding space
is only an artifact for the introduction
Euclidean
canonical commutation rules. It does not belong to
the manifold where the particle moves.
An alternative approach must
therefore be found
where
a memory of the embedding space is absent.

Such an alternative is offered by
gauge theories.
In these,
equations of motion exhibit
symmetries which are local in time
with the consequence
that
the dynamics of some degrees of freedom is
not specified by the equations of motion.
Such local symmetries are gauge symmetries,
and the undetermined degrees of freedom
correspond to pure gauge configurations.
Upon quantization, non-physical degrees of freedom are removed
by restricting the Hilbert space to a physical subspace formed
by gauge invariant states.

Thus, rather than restricting the particle motion
to the surface of a
sphere via constraints as above,
we consider a free motion in a Cartesian
coordinate system,
and impose the condition that
the physical states in Hilbert space are invariant
under arbitrary time-dependent
rescalings of the radial size of the system.
The transverse momentum $\hat{p}_r$
can be made a generator
of gauge transformations,
and we may require $\hat{p}_r^\dagger\hat{p}_r$ to be zero for
physical states.

To achieve this goal, we invoke the method of abelian conversion
that allows one to transform a second-class constrained system into
an abelian gauge theory \cite{fradkin}. Dynamics of physical (gauge
invariant) degrees of freedom in the effective abelian theory is
the same as dynamics of physical degrees of freedom
in the original second-class constrained system.
In general, the abelian conversion
proceeds as follows. Given a set of
second-class constraints $\varphi _a\ (a=1,2,\ldots , 2M$), one extends
the original phase space by extra independent canonical variables
$Q_\alpha,P_\alpha\ (\alpha =1,2,\ldots ,M$).
The extended phase space is equipped
with the canonical symplectic
structure $\{Q_\alpha ,P_\beta\}=\delta _{\alpha \beta}$
and $\{x_i,p_j\}=\delta_{ij}$
(all other
brackets are zero).
With the help of abelian conversion,
quantum dynamics on a manifold can be
formulated independently of the
 parametrization
of the  manifold
\cite{klsh1}.

 An equivalent set of abelian first-class constraints
$\sigma _a=0$
is constructed in such a way that
it
satisfies
\begin{equation}
\{\sigma _a,\sigma _b\}=0.
\label{13}
\end{equation}
This amounts to solving first-order differential equations
with the boundary condition $\sigma _a(x,p,P=0,Q=0)=\varphi _a(x,p)$.
Given the new constraints $\sigma_a=0$, 
the original system Hamiltonian $H(x,p)$
is converted into a Hamiltonian on the extended phase space
$\bar{H}(x,p,P,Q)$ by solving the equation
\begin{equation}
\{\bar{H},\sigma _a\}=0
\label{14}
\end{equation}
with the boundary condition $\bar{H}(x,p,Q=0,P=0)=H(x,p)$.
The extrema of the associated
extended action
${S}^{\rm ext}=\int
dt(p_i\dot{x}_i+P_\alpha\dot{Q}_\alpha -\bar{H}-\lambda _a\sigma _a)$
determine
equations of motion
 in the extended phase space.
These
depend on $2M$ arbitrary functions of time $\lambda _a(t)$
as a manifestation of the gauge
freedom. There exists a choice for $\lambda _a(t)$ such
that the auxiliary phase space variables
$P_\alpha (t)$ and $Q_\alpha(t) $ vanish at all times, whereas $p_i(t)$ and
$x_i(t)$ solve the original equations of motion of the
second-class constrained system \cite{fradkin}.

Applying this procedure
to our particular
second-class constraints (\ref{5}) and (\ref{5p})
yields
\begin{equation}
\sigma _1=\varphi _1+P\ ,\ \ \sigma _2=\varphi _2+2x^2Q\ ,
\label{15}
\end{equation}
while the extended Hamiltonian assumes the form
\begin{equation}
\bar{H}=\frac{1}{2} \left(\frac{\sigma ^2_2}{\sigma _1+R^2}+
\frac{L^2_a}{\sigma _1+R^2}\right)\ ,
\label{16}
\end{equation}
where $L_a=-ipL_ax$ are the classical
components of the angular momentum (for $a=ij$, $L_a=x_ip_j-x_jp_i$).
The extended phase space variables
$p,x,P,Q$ are Cartesian and satisfy the standard Poisson bracket relations.
They can be turned into hermitian operators in the usual way
by the replacement $ \{\ ,\ \}\rightarrow -i[\ ,\ ]/ \hbar$

Physical states are invariant with respect to transformations generated by
$\sigma _{1,2}$:
$x \rightarrow e^{-\xi_2}x$ and $Q\rightarrow Q-\xi _1$, where $\xi _{1,2}$
are parameters of the gauge transformations \cite{klsh1}.
The first is an arbitrary time-dependent rescaling
of the size of the system, which is geometrically
equivalent to the initial restriction
of the motion
to the surface of the sphere, while the second
implies that the auxiliary degree of freedom is a pure gauge.
The first-class constraints
restrict the physical Hilbert space to the gauge-invariant
 sector by the Dirac
conditions \cite{dirac2}
\begin{equation}
\hat{\sigma}_{1,2}\Psi_{\rm phys}=0\ .
\label{17}
\end{equation}
The general solution has the form $\Psi_{\rm phys}=f(Q,x^2)\Psi
(\Omega)$, where $f(x,Q)$ is some fixed function,
whereas $\Psi(\Omega)$ are
wave functions on the $D$-sphere.
In the
physical Hilbert space, we can set
$\sigma _{1,2}$ to zero in the Hamilton operator (\ref{16}).
Thus we find
the energy values
\begin{equation}
E_l=\frac{\hbar ^2}{2R^2}l(l+D-1)\ ,
\label{18}
\end{equation}
rather than (\ref{12}). There is no additional constant energy,
in agreement with  \cite{POD,LL,kleinert}.

This result is
obtained without
ordering problems.
Although ordering ambiguities
are not absent altogether,
they do not affect the final result
 since
they occur only in the quantization
of the gauge generator
$\sigma _2$, where they modify only
the explicit
form of the physically irrelevant function
$f(x^2,Q)$, but not
the physical Hilbert space described by
the wave function $\Psi
(\Omega)$, nor
the
spectrum (\ref{18}).
In fact, in the simple system at hand
the ordering ambiguities in $\hat\sigma_2$
produce only a
multiplicative
renormalization of the physical states
$\Psi_{\rm phys}$.

{\bf 4}. The
above quantization of a particle on a sphere via abelian conversion
is of course applicable to arbitrary homogeneous
spaces.
 But what about arbitrary manifolds?
When attempting a straight-forward
generalization
we run once more into
operator-ordering
problems for the extended Hamiltonian (\ref{16}),
and these require  a new strategy.
The quantum Hamiltonian $\hat{H}= \hat{p}^2/2$
has no ordering ambiguity and  is again adopted as a starting point.
Let $n(x)$ be a unit vector normal to the manifold
in the embedding space
(if this space has a dimension higher than
$D+1$, more normal vectors are needed to specify
transverse directions). The condition $\hat{p}_n \Psi_{\rm phys}
= (n(\hat{x}),\hat{p})\Psi_{\rm phys}= 0$ removes
the transverse
motion and offers itself as a generator of gauge transformations.
This, however, is not consistent
for two reasons.
First,
since $n(\hat{x})$ depends on position,
 $[\hat H, \hat{p}_n]$ is nonzero, so that
the
free-particle  Hamiltonian is not gauge invariant.
Second, the operator
$\hat{p}_n$ is not hermitian making finite gauge transformations non-unitary.

The first problem can be resolved via an abelian conversion method
performed immediately at the quantum level \cite{fradkin}.
Here one uses
{\it operator} versions of second-class constraints
$\hat{\varphi}_1 = F(\hat{x})$ and $\hat{\varphi}_2
=\hat{p}_n$ to restrict
the  motion to a manifold specified by $F(x)=0$,
the forbidden direction being
specified by the normal vector
$n(x) = \partial F(x)/|\partial F(x)|$.
Then one extends the system by two extra canonical
operators $\hat{Q}$ and $\hat{P}$ obeying standard Heisenberg
commutation
relation, and commuting with $\hat{x}$ and $\hat{p}$. The conversion
of the second-class constraint operators $\hat{\varphi}_a$ is
enforced by solving equations (\ref{13}) and (\ref{14}) with
 Poisson brackets replaced by commutators. The abelian
gauge generators have the same basic structure as those in (\ref{15}),
only that
the $\hat x$-dependent factor of $\hat{Q}$ in $\hat{\sigma}_2$ depends
now on $F(\hat x)$. This construction
solves the
first problem of finding a gauge invariant
Hamiltonian operator.

The hermiticity problem for the generator $\hat{\sigma}_2$
must be solved in a way that dynamics of physical degrees
of freedom does not depend on the embedding procedure.
For this we observe that the operator $\hat{\sigma}_2^\dagger\hat{\sigma}_2$
is just as good as $\hat{\sigma}_2$ itself to
eliminate
excitations of the transverse modes.
It has the advantage of being
hermitian. In addition, it
commutes with $\hat{\sigma}_1$, so the abelian gauge
algebra (\ref{13}) is retained by such a choice of the second
generator. Thus we modify the conversion method
by taking
$\hat{p}_n^\dagger\hat{p}_n$ as the constraint
operator $\hat{\varphi}_2$ rather than $\hat{p}_n$.
At the classical level, the new constraint
 $p^2_n=0$ is certainly equivalent to $p_n =0$.
But at the operator level,
it is superior by
generating unitary gauge transformations.
With the new generator,
 the conversion
equation (\ref{14}) gives rise to a new Hamiltonian operator
in the extended Cartesian coordinate system.

It is not hard to verify that the physical states
(\ref{17}) have the form
$e^{iFQ/\hbar}\Psi(x)$ where $\hat{p}_n^\dagger\hat{p}_n \Psi =0$, that is,
the kinetic energy of the transverse motion is strictly zero for
physical states with our choice of the gauge generators.

\end{document}